\documentclass[journal]{rmaa}

\usepackage{paralist}

\usepackage{psfrag,color}

\usepackage[latin1]{inputenc}

\def\gsmf{\mbox{\textit{GSMF}}}
\def\gbmf{\mbox{\textit{GBMF}}}
\def\HMF{\mbox{\textit{HMF}}}
\def\lcdm{\mbox{$\Lambda$CDM}}
\def\ms{\mbox{$M_{\rm s}$}}
\def\mg{\mbox{$M_{\rm g}$}}
\def\mh{\mbox{$M_{\rm h}$}}
\def\mb{\mbox{$M_{\rm b}$}}
\def\mtran{\mbox{$M_{\rm tran}$}}
\def\mcross{\mbox{$M_{\rm cross}$}}
\def\mpch{\mbox{$h^{-1}$Mpc}}
\def\msunh{\mbox{$h^{-1}$M$_\odot$}}    
\def\fs{\mbox{$f_{\rm s}$}}
\def\fg{\mbox{$f_{\rm g}$}}
\def\fb{\mbox{$f_{\rm b}$}}

\def\ltsima{$\; \buildrel < \over \sim \;$}    
\def\lesssim{\lower.5ex\hbox{\ltsima}}           
\def\gtsima{$\; \buildrel > \over \sim \;$}    
\def\grtsim{\lower.5ex\hbox{\gtsima}}           

\def\msun{\mbox{M$_{\odot}$}}

\def\nyu{\mbox{NYU-VAGC}}
\def\mnras{\mbox{MNRAS}}
\def\apj{\mbox{ApJ}}
\def\aap{\mbox{A\&A}}
\def\apjs{\mbox{ApJS}}
\def\aj{\mbox{AJ}}
\def\pasp{\mbox{PASP}}
\def\jcap{\mbox{JCAP}}

\title{On the stellar and baryonic mass fractions of central blue and red galaxies}

\author{
A. Rodr\'iguez-Puebla,\altaffilmark{1} 
V. Avila-Reese,\altaffilmark{1}
C. Firmani,\altaffilmark{1,2}
P. Col\'{\i}n,\altaffilmark{3}}

\altaffiltext{1}{Instituto de Astronom\'ia, Universidad Nacional Aut\'onoma de M\'exico. Email: apuebla@astroscu.unam.mx.}
\altaffiltext{2}{Osservatorio Astronomico di Brera, Italy.}
\altaffiltext{3}{Centro de Radioastronom\'{\i}a y Astrof\'{\i}sica, Universidad Nacional 
Aut\'onoma de M\'exico.}
\shortauthor{Rodr\'iguez-Puebla et al.}
\shorttitle{Stellar/baryonic mass fractions of blue and red galaxies}

\fulladdresses{
\item A. Rodr\'iguez-Puebla \& V. Avila-Reese: 
Instituto de Astronom\'ia, Universidad Nacional Aut\'onoma de M\'exico, 
A.P. 70-264, 04510, M\'exico, D.F., M\'exico.
\item C. Firmani: Osservatorio Astronomico di Brera, via E.Bianchi 46, I-23807 Merate, Italy.
\item P. Col\'{\i}n: Centro de Radioastronom\'{\i}a y Astrof\'{\i}sica, Universidad Nacional 
Aut\'onoma de M\'exico, A.P. 72-3 (Xangari), Morelia, Michoac\'an 58089, M\'exico.}

\listofauthors{A. Rodr\'iguez-Puebla, V. Avila-Reese, C. Firmani \& P. Col\'{\i}n}
\indexauthor{Rodr\'iguez-Puebla, A.}
\indexauthor{Avila-Reese, V.}
\indexauthor{Firmani, C.}
\indexauthor{Col\'{\i}n, P.}

\abstract{Using the abundance matching technique, we infer the local stellar and 
baryonic mass--halo mass (\ms-\mh\ and \mb-\mh) relations separately for {\it central} blue and red 
galaxies (BGs and RGs). The observational inputs are the SDSS central BG and RG Stellar Mass 
Functions and the measured gas mass-\ms\ relations. For halos associated to central BGs, 
the distinct \lcdm\ Halo Mass Function is used and set up to exclude: (i) the observed 
group/cluster mass function and (ii) halos with a central major merger at resdshifts $z\le0.8$. For 
central RGs, the complement of this mass function to the total one is used. At $\mh>10^{11.5} \msun$, 
the \ms\ of RGs tend to be higher than those of BGs for a given \mh, the difference not being larger 
than 1.7. At $\mh<10^{11.5} \msun$, this trend is inverted. For BGs (RGs): (a) the maximum value 
of $\fs=\ms/\mh$ is $0.021^{+0.016}_{-0.009}$ ($0.034^{+0.026}_{-0.015}$) and it is attained at 
$\log$(\mh/\msun)$=12.0$ (=11.9);  (b) $\fs\propto$ \mh\ ($\fs\propto \mh^3$) at the low-mass end while at 
the high-mass end, $\fs\propto \mh^{-0.4}$ ($\fs\propto \mh^{-0.6}$). The baryon mass fractions, 
\fb=\mb/\mh, of BGs and RGs reach maximum values of $\fb=0.028^{+0.018}_{-0.011}$ and 
$\fb=0.034^{+0.025}_{-0.014}$, respectively. At $\mh<10^{11.3} \msun$, the dependence of 
\fb\ on \mh\ is much steeper for RGs than for BGs. We discuss on the differences 
found in the \fs-\mh\ and \fb-\mh\ relations between BGs and RGs in the light of semi-empirical 
galaxy evolution inferences.}

\resumen{
}

\addkeyword{cosmology: dark matter}
\addkeyword{galaxies: mass function}

\begin{document}
\maketitle

\section{Introduction}

The galaxy stellar and baryonic mass functions (\gsmf\ and \gbmf, respectively),
inferred from the observed luminosity function and gas fraction--stellar mass (\fg--\ms) relation, 
contain key statistical information to understand the physical processes 
of galaxy formation and evolution. Within the context of the popular $\Lambda$ Cold Dark Matter (\lcdm) 
hierarchical scenario, dark matter halos are the sites where galaxies form and evolve (White \& Rees
1978; White \& Frenk 1991). 
Hence, a connection between \gbmf\ or \gsmf\ and the halo 
mass function (\HMF) is expected.  The result of such a connection is the galaxy stellar and baryonic 
mass--halo mass relations, \ms-\mh\ and \mb--\mh, and their intrinsic scatters, both set
by complex dynamical  and astrophysical processes intervening in galaxy formation and 
evolution (see for recent reviews Baugh 2006; Avila-Reese 2007; Benson 2010). In this sense, 
the \mb/\mh\ and \ms/\mh\ ratios quantify the efficiency at 
which galaxy and star formation proceeds within a halo of mass \mh. Therefore, the empirical 
or semi-empirical inference of the \mb--\mh\ and \ms--\mh\ relations and their scatters 
(locally and at other epochs) is nowadays a challenge of great relevance in astronomy.

For simplicity, in statistical studies like those related to the \gsmf, galaxies
are labelled by their mass alone. However, by their observed 
properties, correlations, and evolution, galaxies show a very different nature, 
at least for the two major groups in which they are classified: the rotationally-supported 
disk star-forming (late-type) and the pressure-supported spheroid quiescent (early-type).
In the same way, the evolution of galaxies is expected to differ if they are centrals or satellites.
The main intrinsic processes of galaxy evolution are associated
to central galaxies, while satellite galaxies undergo several {\it extra} astrophysical processes because of 
the influence of the environment of the central galaxy/halo system in which they were accreted.
Hence, if the \mb--\mh\ or \ms--\mh\ relations are used for constraining galaxy formation and 
evolution processes, these relations are required by separate for at least the two main 
families of late- and early-type galaxies and taking into account whether the galaxy is central 
or satellite. Fortunately, in the last years there appeared 
several studies, in which a decomposition of complete {\gsmf}s by color, concentration
or other easily measurable indicators of the galaxy type was carried out (e.g., Bell et al. 2003; 
Shao et al. 2007; Bernardi et al. 2010). Evenmore, in a recent work, Yang, Mo \& van den Bosch 
(2009, hereafter YMB09) 
used the Sloan Digital Sky Survey (SDSS) data for obtaining the {\gsmf}s of both central and 
central + satellite galaxies separated in each case into blue and red objects.

With the coming of large galaxy surveys, a big effort has been done in 
constraining the $z\sim 0$ {\it total} \ms--\mh\ relation (i) {\it directly} by estimating halo masses with 
galaxy-galaxy weak lensing, with kinematics of satellite galaxies or with X-ray studies; and (ii) 
{\it indirectly} by linking observed statistical galaxy properties (e.g., the galaxy stellar mass function \gsmf, 
the two-point correlation function, galaxy group catalogs) to the theoretical \HMF\ (see for recent 
reviews and more references Moster et al. 2010; Behroozi, Conroy \& Wechsler 2010, hereafter
BCW10; More et al. 2011). 
While the latter approach does not imply a measure-based determination of halo masses, it is simpler from 
a practical point of view, as it allows to cover larger mass ranges, and can be extended to higher redshifts than the 
former approach (see recent results in Conroy \& Wechsler 2009; Moster et al. 2010; Wang \& Jing 2010; 
BCW10). Besides, both the weak lensing and satellite kinematics methods in practice 
are (still) statistical in the sense that one needs to stack large number of galaxies in order to get 
sufficient signal-to-noise. This introduces a significant statistical uncertainty in the inferred halo masses. 

The indirect approach for linking galaxy and halo masses spans a large
variety of methods, among them the Halo Occupation Distribution (Peacock \& Smith 2000; 
Berlind \& Weinberg 2002; Kravtsov et al. 2004) and the Conditional Luminosity Function formalisms (Yang et al.
2003, 2004). These formalisms introduce a priori functional forms with several parameters
 that should be constrained by the observations. Therefore, the final inferred \ms--\mh\ relation is 
actually model-dependent and yet sometimes poorly constrained due to degeneracies in the
large number of parameters.  A simpler and more
empirical method --in the sense that it uses only the \gsmf\ (or luminosity function) as input 
and does not require to introduce any model-- has been found to give reasonable results. This 
indirect method, called the abundance matching technique (hereafter AMT; e.g., Marinoni \& Hudson 2002; 
Vale \& Ostriker 2004; Conroy et al. 2006; Shankar et al. 2006; Conroy \& Wechsler 
2009; Baldry, Glazebrook \& Driver 2008; Guo et al. 2010; Moster et al. 2010; BCW10), 
is based on the assumption of a monotonic correspondence between \ms\ and \mh; in the 
limit of zero scatter in the \ms--\mh\ relation, the halo mass \mh\ corresponding to a galaxy of 
stellar mass \ms, is found by matching the observed cumulative \gsmf\ to the theoretical cumulative \HMF.  
 
In this paper we apply the AMT in order to infer the local 
\ms--\mh\ relation for {\it central blue and red} galaxies 
separately, which requires as input \textit{both} the observed 
central blue and red {\gsmf}s, taken here from YMB09.  
Note that in order to infer the \ms--\mh\ relation of galaxy subpopulations
(e.g., blue/red or central/satellite ones) solely from the overall 
\gsmf, models for each subpopulation should be introduced, 
which largely increases the uncertainty in the result.
Regarding the {\HMF}s to be matched
with the corresponding observed central {\gsmf}s, 
the theoretical \HMF\ is decomposed into two functions
--associated to halos hosting blue and red galaxies-- 
based on empirical facts:
blue galaxies are rare as central objects 
in groups/clusters of galaxies, and they should not have undergone late major mergers 
because of the dynamical fragility of disk (blue) galaxies. Nowadays, it is not clear whether the \ms--\mh\ 
relation varies significantly or not with galaxy color or type. Previous studies that discussed this 
question were based on direct methods: the weak lensing (Mandelbaum et al. 2006) and 
satellite kinematics (More et al. 2011) techniques. The uncertainties in the results of 
these studies are yet large, and can be subject to biases intrinsic to the sample selection 
and to effects of environment. 

We also estimate here the galaxy baryon mass-halo mass relations, 
\mb--\mh\footnote{We assume that the galaxy baryonic mass is included in the 
halo (virial) mass \mh.}, where \mb = \ms + \mg, by using the {\gsmf}s combined with average 
observational determinations of the galaxy gas mass, \mg, as a function of \ms.
The galaxy baryonic mass fraction, \fb = \mb/\mh, and its dependence on mass 
is important for constraining models and simulations of galaxy evolution, and is also 
a key input for some approaches, implemented for modelling the most 
generic population of galaxies, namely {\it isolated (central) disk galaxies} 
(e.g., Mo, Mao \& White 1998; Firmani \& Avila-Reese 2000; van den Bosch 2000; 
Stringer \& Benson 2007; Dutton et al. 2007; Gnedin et al. 2007; Dutton \& van den 
Bosch 2009).  In these and other studies, it was shown that several disk galaxy properties, 
correlations and their scatters depend (or are constrained) by \fb. In a similar way, 
the \fb\--\mh\ dependence is expected to play some role in the results of structural 
and dynamical models of spheroid-dominated galaxies.

In Section 2 we describe the method and the data input. The stellar/baryon mass--halo mass 
relations for the total, blue and red (sub)samples are presented in Section 3. In Section 4 we 
compare our results with other observational works, and discuss whether they
are consistent or not with expectations of semi-empirical inferences.
The summary and our conclusions are given in Section 5.

\section{The method}

The AM statistical technique is based on the hypothesis of a one-to-one monotonic 
increasing relationship between \ms\ (or \mb) and \mh. Therefore, by matching
the {\it cumulative} galaxy stellar and halo mass functions, for a given \ms\ a 
unique \mh\ is assigned:
\begin{equation}
\int _{\mh} ^{\infty} \phi_{h} (M_{h}^{\prime})dM_{h}^{\prime}=\int _{\ms} ^{\infty} \phi_{s}(M_{s}^{\prime}) dM_{s}^{\prime},
\end{equation}
where $\phi_h$ is the overall \HMF\ (distinct + subhalos) and $\phi_s$ is the overall \gsmf;
\textit{distinct} halos are those not contained inside more massive halos. 
It is reasonable to link central galaxies with distinct halos. Therefore, in the case of using
the \gsmf\ for only central galaxies, the distinct \HMF\ should be used for the matching. 
Since the main purpose of this paper is 
the inference of the \ms--\mh\ (and the corresponding \mb--\mh) relation for blue (red) galaxies, 
(i) a \gsmf\ that separates galaxies by color is necessary (the data to be used here are discussed 
in \S\S 2.1), and (ii) a criterion to select the halos that will likely host blue (red) 
galaxies shall be introduced (see \S\S 2.2.1).

In this paper we will not carry out an exhaustive analysis of uncertainties in the 
inference of the \ms--\mh\ relation with the AMT. This was extensively done 
in BCW10 (see also Moster 
et al. 2009). In BCW10 the uncertainty sources are separated into three classes: uncertainties 
(i) in the observational inference of \gsmf, (ii) in the dark matter HMF, which includes uncertainties 
in the cosmological parameters, and (iii) in the matching process arising primarily from the intrinsic 
scatter between \ms\ and \mh. 

\subsection{Galaxy and Baryonic Stellar Mass Functions}

\begin{figure*}
\vspace{8.2cm}
\hspace{-1.0cm}
\includegraphics{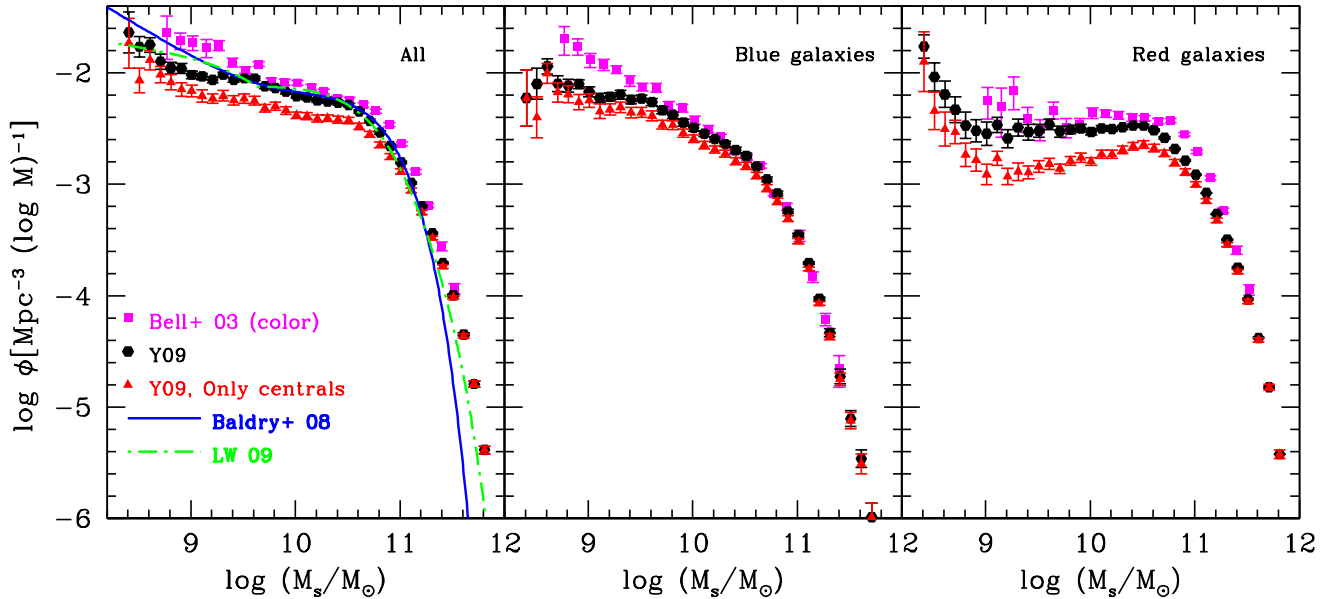}
\caption{ \textit{Left panel:}  Different local {\gsmf}s for {\it all} 
galaxies. The reported data in Bell et al. (2003, pink squares) and YMB09 (black hexagons)
are plotted directly, while for Baldry et al. (2008, blue solid line) and Li \& White (2009, dot-dashed green line),
the best fits these authors find to their samples are plotted. Red triangles show the data from YMB09
corresponding to the \gsmf\ of central-only galaxies.
\textit{Middle and left panels:} Data corresponding to the decomposition
of the \gsmf\ into blue and red galaxies, respectively, from Bell et al. (2003) and 
for the all and central-only galaxies from YMB09.
}
\label{GSMFs}
\end{figure*}
In the last years, complete galaxy luminosity
functions (and therefore, {\gsmf}s) were determined for local samples covering a large range 
of luminosities (masses). The stellar mass is inferred from (multi)photometric and/or spectral 
data (i) by using average stellar mass-to-light ratios, depending only on color (inferred from 
application of stellar population synthesis --SPS-- models to galaxy samples with independent 
mass estimates, e.g. Bell et al. 2003), or (ii) by applying directly the SPS technique to 
each sample galaxy, when extensive multi-wavelength and/or spectral information is available. 

In both cases, a large uncertainty is introduced in the inference of \ms\ due to the 
uncertainties in the IMF, stellar evolution, stellar spectral libraries, dust extinction, metallicity, 
etc. Bell et al. (2003) estimated a scatter of $\approx 0.1$ dex in their \ms/$L$ ratios in infrared 
bands. Conroy, Gunn \& White (2009) carried out a deep analysis of propagation of uncertainties 
in SPS modelling and concluded that \ms\ at $z\sim 0$ carry errors up to $\sim 0.3$ dex 
(but see Gallazzi \& Bell(2009)). Here, we will consider an overall systematical uncertainty 
of 0.25 dex in the \ms\ determination (see BCW10). 

Most of the current local {\gsmf}s were inferred from 2dF Galaxy Redshift Survey, 
Two Micron All-Sky Survey (2MASS) and SDSS (e.g., Cole et al. 2001; Bell et al. 2003;
Baldry et al. 2006). The low-mass completeness limit due to missing of low 
surface brightness galaxies is at $\sim 10^{8.5}$ \ms\ (Baldry et al. 2008). An
upturn of the \gsmf\ close to this end (below \ms$\sim 10^9$ \msun) was confirmed in 
several recent works (Baldry et al. 2008; YMB09; Li \& White 2009). Due to this upturn, 
a better fit to the {\gsmf}s is obtained by using  a double or even triple Schechter 
function. Since the low-mass end of the \gsmf\ is dominated by late-type galaxies,
this upturn plays an important role in the \ms--\mh\ relation of late-type 
galaxies at low masses.

For our purposes, observational works where the \gsmf\ is decomposed into late- 
and early-types galaxies are required. Such a decomposition has been done, for example, 
in Bell et al (2003), who combined 22679 SDSS Early Data Release and 2MASS galaxies, 
and used two different criteria, color and concentration, to split the sample into two 
types of galaxies. A much larger sample taken from the {\nyu} based on the SDSS DR4 
has been used by YMB09 (see also Yang, Mo \& van den Bosch 2008), who split
the sample into {\it blue and red} subsamples according to a criterion in the 
$^{0.1}(g-r)-M_r$ diagram.
In both works, \ms\ is calculated from the $r-$band magnitude 
by using the corresponding color-dependent \ms/$L_r$ ratio given in Bell et al. 
(2003). In YMB09 each color subsample is in turn separated into central and satellite 
galaxies according to their memberships in the constructed groups, where the central galaxy
is defined as the most massive one in the group and the rest as satellite galaxies.

In Figure \ref{GSMFs}, the Bell et al. (2003) and YMB09 {\gsmf}s are reproduced by using 
the data sets reported in these papers. In the left panel, the full sample from each work 
(solid squares and solid hexagons, respectively) are plotted, as well as the case of 
central-only galaxies from YMB09 (solid triangles); both {\gsmf}s and the other ones
plotted in this figure are normalised to $h=0.7$ and to a Chabrier (2003) IMF. 
In the central and right panels, the corresponding blue (late-type) and red (early-type) 
sub-samples are plotted with the same symbols on the left panel. For the Bell et al. 
sub-samples, only those separated by their color criterion are plotted. Both {\gsmf}s 
corresponding to the full and blue sub-samples are in good agreement for 
$\ms\grtsim 10^{9.5}$ \msun. For lower masses, the Bell et al. \gsmf's are higher.
On one hand, the Bell et al. sample is much smaller than the YMB09 one (therefore its cosmic 
variance is more significant). On the other hand, the redshift completeness and \ms\ limit 
in YMB09 is treated with updated criteria.

In Figure \ref{GSMFs}, we also plot fits to the overall \gsmf\ presented in Baldry et al. (2008, 
double Schechter function, solid blue line) and in Li \& White (2009, triple 
Schechter function, dashed green line) for new SDSS releases and by using 
directly SPS models to estimate \ms\ for each galaxy. These fits agree well with the YMB09 data 
in the mass range $9.2\lesssim$ log(\ms/\msun) $\lesssim 11.2$. For smaller masses, the Baldry 
et al. fit tends to be steeper while the Li \& White fit tends to be shallower than the YMB09 data.  
For larger masses, both fits decrease faster with \ms\ than the YMB09 data. All these (small) 
differences are due to the different methods used to estimate \ms, as well as the different volumes 
and limit corrections of the samples (see Baldry et al. 2008, YMB09, and Li \& White 2009 for discussions). 

The split into two colors of the sample used by YMB09 is a rough approximation
to the two main families of disk- and spheroid-dominated galaxies. It is well known 
that the morphological type correlates with the galaxy color, though with a large scatter. 
There is for example a non-negligible fraction of galaxies (mostly highly inclined) that are 
red but of disk-like type (e.g., Bernardi et al. 2010). However, given that here we consider 
a partition of the overall sample just in two groups, we believe that is reasonable to assume 
in a first approximation that the color criterion for the partition will provide a similar result at 
this level as a morphological criterion. 

\begin{figure}
\vspace{6.3cm}
\includegraphics{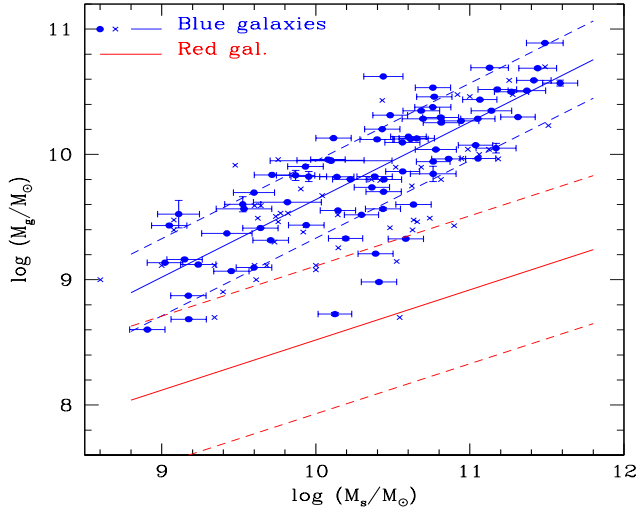}
\caption{Gas mass vs stellar mass for a sample of disk high and low surface brightness galaxies collected and
homogenised by Avila-Reese et al. (2008, blue dots with error bars) and for a sample of disk
galaxies presented by McGaugh (2005, blue crosses). The solid blue line is the orthogonal linear doubly-weighted
regression to the data from the former authors and the dashed lines show an estimate of the
intrinsic scatter around the fit. The solid red line is an estimate of the \mg--\ms\ correlation for
red galaxies by using our fit to blue galaxies and the ratio of blue-to-red atomic gas fraction 
determined in Wei et al. (2010; see text). }
\label{gas}
\end{figure}

For the YMB09 sample, the blue and red galaxies are $\approx 55\%$ and $\approx 45\%$, 
respectively, for $\ms\grtsim 3\times 10^8$ \msun.  Red galaxies dominate the total \gsmf\ at large
masses. At \ms$\approx 2\times 10^{10}$ \msun\ the abundances of red and blue galaxies are similar
and at lower masses the latter are increasingly more abundant than the former as \ms\ is smaller. 
For $\ms\lesssim 10^9$ \msun, the abundance of red galaxies, mainly central ones, steeply increases 
towards smaller masses. The existence of this peculiar population of faint central red galaxies 
is discussed in YMB09. Wang et al. (2009) suggested that these galaxies
are hosted by small halos that have passed through their massive neighbors, and the 
same environmental effects that cause satellite galaxies to become red are also responsible
for the red colors of such galaxies.  However, as these authors showed, even if the
environmental effects work, there are in any case over 30\% of small halos that are
completely isolated in such a way that these effects can not be invoked for them. 

In the YMB09 sample, around $70$\% of the galaxies are central. As mentioned in the Introduction, 
the inference of the \ms--\mh\ relation for central-only galaxies is important for studies aimed
to constrain galaxy formation and evolution in general; satellite galaxies are interesting on its own 
but they lack generality because their evolution and properties are affected by extra environmental 
processes.  

In what follows, the YMB09 \gsmf\ provided in tabular form and split into blue/red and central/satellite
galaxies will be used for applying the AMT. Our main goal is to infer the \ms--\mh\ relation
for central blue (late-type) and red (early-type) galaxies. 

We will infer also the corresponding \mb--\mh\ (baryonic) relations. 
The blue and red {\gbmf}s are estimated from the blue and red {\gsmf}s, respectively, 
where in order to pass from \ms\ to \mb, the cool (atomic and molecular) gas mass, \mg, 
corresponding on average to a given \ms\ is taken from the empirical blue and
red \mg--\ms\ relations. In Figure \ref{gas}, a compilation of 
observational estimates is plotted in the \ms--\mg\ plane for a sample of disk galaxies 
that includes low surface brightness galaxies from Avila-Reese et al. (2008; blue 
dots with error bars; they added $H_2$ mass contribution by using an estimate
for the $H_2$-to-$HI$ mass ratio as a function of galaxy type), and 
for another galaxy sample from McGaugh (2005; blue crosses; no $H_2$ contribution
is considered and their dwarf galaxies were excluded).  
An orthogonal linear doubly-weighted regression to the data from Avila-Reese et al. 
(2008) gives:
\begin{equation}
\frac{\mg}{10^{10}\msun} = 0.43\times \left(\frac{\ms}{10^{10}\msun}\right)^{0.62}.
\label{mgas}
\end{equation}
This fit is plotted in Figure \ref{gas} with its corresponding estimated
scatter ($\approx 0.3$ dex; blue solid and dashed lines). This is the relation and 
its scatter used to calculate \mb\ and the blue \gbmf. A similar relation has been 
inferred by Stewart et al. (2009). The gas fractions in red galaxies are much 
smaller than in the blue galaxies. For sub-samples of blue and red galaxies, 
Wei et al. (2010) reported for each one the atomic gas fractions versus \ms\
(molecular gas was not included). The ratio of their fits to these data as a function 
of \ms\ 
is used here to estimate from eq. (\ref{mgas}, blue galaxies) the corresponding average \mg\ for 
red galaxies as a function of \ms. The red solid line shows the obtained relationship. 
As an approximation to the scatter (short-dashed lines), the average scatter reported
for red galaxies in Wei et al. (2010) is adopted here.

\subsection{Halo and sub-halo mass functions}

A great effort has been done in the last decade to determine the \HMF\ at $z=0$ and at higher 
redshifts in N-body cosmological simulations. A good fit to the results, at least for low
redshifts, is the universal function derived from a Press-Schechter formalism (Press \& Schechter 1974) 
generalized to the elliptical gravitational collapse (Sheth \& Tormen 1999 hereafter S-T). In fact, Tinker 
et al. (2008) have shown that at the level of high precision, the \HMF\ 
may change for different cosmological models and halo mass definitions
as well as a function of $z$. For our purposes and for the cosmology used 
here, the S-T approximation provides a good description of the $z=0$ 
\HMF\ of distinct halos:
\begin{eqnarray}
\phi_h(M_h)dM_h &=& \nonumber \\
A\left(1+\frac{1}{\nu^{2q}}\right)\sqrt{\frac{2}{\pi}}\frac{\bar{\rho}_{M}\nu}{M_h^2}\left|\frac{d\ln \sigma}{d \ln M_h}\right| \exp\left[-\frac{\nu^2}{2}\right]dM_h
\label{ST}
\end{eqnarray} 
where $A= 0.322$, $q=0.3$, $\nu^2=a(\delta_c/D(z)\sigma(\mh))$  with $a=0.707$, 
$\delta_c=1.686\Omega_m^{0.0055}$ is the linear threshold in the case for spherical collapse 
in a flat universe with cosmological constant, $D(z)$ is the growth factor and $\sigma(\mh)$ 
is the mass power spectrum variance of fluctuations linearly extrapolated to $z=0$. 
The halo (virial) mass, \mh\, is defined 
in this paper as the mass enclosed within the radius where the overdensity is $\bar{\rho}_{\rm vir}=\Delta$
times the {\it mean matter} density, $\bar{\rho}_{M}$; $\Delta\approx 340$ according to 
the spherical collapse model for the cosmology used here. 
The cosmological parameters assumed here are close to those of WMAP5 (Komatsu et al. 2009):
$\Omega_M=0.27, \Omega_\Lambda=1-\Omega_m=0.73, h=0.70, \sigma_8=0.8$. 

The distinct \HMF\ should be corrected when a \gsmf\ corresponding to {\it all} galaxies is 
used in the AMT. In this case, satellite galaxies are included in the \gsmf. Therefore, subhalos 
should be taken into account in the \HMF. The subhalo fraction is not more than $\approx 20\%$ 
of all the halos at $z=0$ (e.g., Shankar et al. 2006; Conroy et al. 2006; Giocoli et al. 2010; 
BCW10).  When necessary, we correct the S-T \HMF\ for (present-day) subhalo 
population by using the fitting formula to numerical results given in Giaccoli et al. (2010):
\begin{equation}
\frac{dn(m_{sub})}{d\ln m_{sub}}=A_0m^{\eta-1}_{sub}\exp\left[-\left(\frac{m_{sub}}{m_0}\right)^\gamma\right],
\end{equation}
with $\eta=0.07930$, $\log A_0=7.812$, $\log (m_0/\msun)=13.10$ and $\gamma=0.407$.

\begin{figure}
\vspace{13.4cm}
\includegraphics{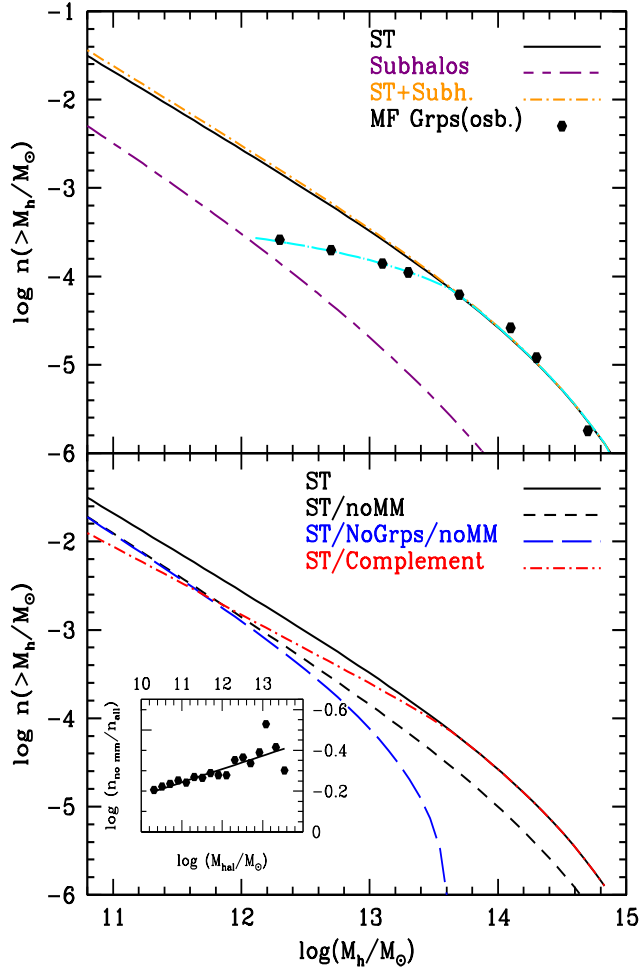}
\caption{\textit{Upper panel:} Distinct S-T \HMF\ for the cosmology adopted in this paper 
(solid black line), sub-halo mass function at $z=0$ according to Giaccoli et al. (2010,
short-long-dashed purple line), and the sum of both (dot-dashed orange line).
The solid dots are measures of the group/cluster mass function according to
Hein{\"a}m{\"a}ki et al. (2003) and adequately corrected to our definition of virial halo mass;
the dot-long-dashed cyan line is a eye-fit to the data.
\textit{Lower panel:} The same distinct S-T \HMF\ (solid black line) shown in the upper panel 
but (i) excluding the halos that suffered late major mergers --since $z=0.8$-- (short-dashed
black line) and (ii) excluding these halos and those of observed groups/clusters (long-dashed
blue line). The latter is the \HMF\ to be assigned to the sub-sample of central blue galaxies. The 
complement of this function to the total (S-T) one (dot-dashed red line) is the \HMF\ to be
assigned to the sub-sample of central red galaxies. The inset shows the ratio of number densities of halos that did 
not suffer major mergers since $z=0.8$ to all the (distinct) halos according to measures in a cosmological
N-body simulation (Col\'in et al. 2011, see text).  The fit to this ratio (solid line in the inset)
is what has been used to correct the S-T \HMF\ for halos thad did not suffer late major mergers.}
\label{HMFs}
\end{figure}

The upper panel of Figure \ref{HMFs} shows the (distinct) S-T \HMF\ (solid line),
the sub-halo \HMF\ (short-long-dashed line), and the distinct+subhalo \HMF\ (dash-dotted 
line). The correction by sub-halos in the abundance is small at low masses and negligible 
at high masses.  When the \gsmf\ refers only to central galaxies --which is the case
in this paper--, then it is adequate to use namely the distinct \HMF\ for 
the AMT, i.e. {\it the subhalo abundance correction is not necessary}.

\subsubsection{Haloes hosting blue and red galaxies}

In the AMT, the cumulative \gsmf\ and \HMF\ are matched in order to link a given \ms\ to \mh.
When a subsample of the total \gsmf\ is used --as is the case for inferring the \ms--\mh\
relation of only late- or early-type galaxies-- it would not be correct to use the total \HMF\ for
the matching. This function, in the ignorance of which is the mass function of halos hosting blue (red) 
galaxies, at least should be re-normalised (decreased uniformly) by the same fraction
corresponding to the decrease of the sub-sample \gsmf\ with respect to the total \gsmf. In YMB09,  
$\approx 55\%$ ($\approx 45\%$) of the galaxies are in the blue (red) sub-samples for
$\ms\grtsim 3\times 10^8$ \msun. We may go one step further by proposing general 
observational/physical conditions for halos to be the host of blue 
(late-type) or red (early-type) galaxies. Note that the division we do here of galaxies
is quite broad --just in two groups--, therefore very general conditions are enough.  

Haloes that host central blue and red galaxies are expected to have (i) a different environment, 
and (ii) a different merger history. We take into account these two factors in order to roughly
estimate the \HMF\ of those halos that will host today central blue and red galaxies.

{\it Environment.-} Blue (late-type) galaxies are rare in the centers of groups and clusters of 
galaxies (high-density environments; e.g., Norberg et al. 2001;  Zehavi et al. 2005; 
Li et al. 2006; de Lapparent \& Slezak 2007; Padilla, Lambas \&  Gonz{\'a}lez 2010; 
Blanton \& Moustakas 2009, and more references
therein). For example, in the SDSS YMB09 sample that
we use here (see also Weinmann et al. 2006), among the groups with 3 or more members, 
the fraction of those with a central blue galaxy is only $\approx 20\%$, and most of these
central galaxies are actually of low masses.  Therefore, cluster- and group-sized halos 
(more massive than a given mass) can not be associated to central blue galaxies when 
using the AMT. This means that the halo mass function of groups/clusters of 
galaxies should be excluded from the theoretical \HMF\ (Shankar et al. 2006). 

Hein{\"a}m{\"a}ki et al. (2003) have determined the \HMF\ of groups with 3 or 
more members and with a number density enhancement $\delta n/n\ge80$ from
the Las Campanas Redshift Survey. The authors estimated the corresponding group virial mass on 
the basis of the line-of-sight velocity and harmonic radius of the group, in such a way that
this mass is defined at the radius where $\delta n/n$=80. The observational galaxy overdensity
$\delta n/n$ is related to the mass overdensity $\delta\rho/\rho$ roughly through the bias parameter $b$:
 $\delta\rho/\rho=$($1/b$)$\times\delta N/N$, where $b\approx 1/\sigma_8$ (Mart{\'{\i}}nez et al. 2002). 
 Hence, for $\sigma_8=0.8$, $\delta\rho/\rho\approx 64$; since the group selection was carried 
 out in Tucker et al. (2000), where
 an Einstein-de Sitter cosmology was used, then $\rho=\rho_{\rm crit}$ in this case. In our case, 
 the halo virial mass is defined at the radius where $\delta\rho/\rho\approx 340$ (see \S\S 2.2); in 
 terms of $\rho_{\rm crit}$, our overdensity is $340\times \Omega_M=92$. Therefore, the halo virial 
masses in Hein{\"a}m{\"a}ki et al. (2003) should be slightly larger than those used here. For the 
NFW halos of masses larger than $\sim 10^{13}$ \msun, the differences are estimated to be 
factors 1.10-1.20.  We  correct the group masses of Hein{\"a}m{\"a}ki et al. (2003) by 15\%. 
In the upper panel of Fig. \ref{HMFs}, the corrected group (halo) mass function is reproduced
(solid dots) and a eye-fit to them is plotted (dot-dashed cyan line). 

{\it Merger history.-}  Disk (blue, late-type) galaxies are dynamically fragile systems and thus they 
are not expected to survive strong perturbations such as those produced in major mergers or 
close interactions. However, as several theoretical studies have shown (e.g., Robertson et al.
2004; Governato et al. 2008), when the mergers are gas-rich ('wet') and/or at early epochs 
(in fact, both facts are expected to be correlated), it is highly probable that a gaseous disk is
regenerated or formed again with the late accreted gas. Therefore, a reasonable restriction
for halos that will host disk galaxies is that they did not undergo {\it central} major mergers
since a given epoch (at earlier epochs, while the central major merger may destroy the
disk, a new gaseous disk can be formed later on). Based on numerical simulations, Governato 
et al. (2008) suggested that a 'wet' major merger of disk galaxies at $z\sim 0.8$ has yet a 
non-negligible probability of rebuilding a significant disk by $z\sim 0$. We will assume here 
that halos for which their {\it centers} have a major merger at $z<0.8$ will not host a disk galaxy.

In Col\'{\i}n et al. (2011; in prep.) the present-day abundance fraction of halos with no {\it central} major merger since
$z=0.8$ was measured as a function of \mh\ from an N-body \lcdm\ cosmological high--resolution simulation
with $\Omega_m = 0.24$, $\Omega_\Lambda = 0.76$, and $\sigma_8 = 0.75$ (box size and
mass per particle of 64 \mpch\ and $1.64 \times 10^7 \msunh$, respectively). The friends-of-friends (FOF)
method with a linking-length parameter of 0.17 was applied for identifying halos. The mass ratio 
to define a major merger was $q=M_{\rm h,2}/M_{\rm h,1}>0.2$ and the merger epoch was estimated as that one when 
the center of the accreted halo arrived to the center of the larger halo by dynamical 
friction; this epoch is calculated as the cosmic time when both FOF halos have "touched" 
plus the respective dynamical friction (merging) time as given by the approximation of 
Boylan-Kolchin et al. (2008). The fraction of halos that did not suffer a major merger since 
$z=0.8$ with respect to all the halos as a function of \mh\ measured in Col\'{\i}n 
et al. (2011) is used here to correct our distinct S-T \HMF. This measured fraction is showed 
in the inset in the lower panel of Fig. \ref{HMFs}; the solid line is a linear fit by eye in the 
log-log plot: log($n_{\rm no MM}/n_{\rm all}$)= $0.472 - 0.065$log(\mh/\msun).
As it is seen, the fraction slightly decreases with mass, which is consistent with the idea that
larger mass halos are assembling later with a significant fraction of their masses being acquired in late 
major mergers. After the correction mentioned above, we get the mass function of halos that did not suffer a central 
major merger ($q>0.2$) since $z=0.8$ (short-dashed black line in the lower panel of Fig. \ref{HMFs}). 
 
 {\it The final corrected {\HMF}s.-}
 The function obtained after (i) subtracting from the distinct S-T \HMF\ the group mass function and (ii) excluding 
 halos that did not suffer a late central major merger is plotted in Fig. \ref{HMFs} (blue long-dashed line).  
 This mass function is proposed here to correspond to halos that host today
 blue galaxies. The overall number fraction of these halos with respect to the distinct ones (described
 by the S-T \HMF) is $\sim 58\%$, which is roughly consistent with the fraction of blue galaxies 
 in the YMB09 sample. The \HMF\ corresponding to the complement is plotted in Fig. \ref{HMFs} as
 the red dot-dashed curve. By exclusion, this \HMF\ will be associated with the \gsmf\ of the 
 red central galaxy sub-sample for deriving the \ms--\mh\ relation of red galaxies.
 
\section{Results}

\subsection{The overall, central, and satellite stellar--halo mass relations} 

\begin{figure}
\vspace{14.4cm}
\includegraphics{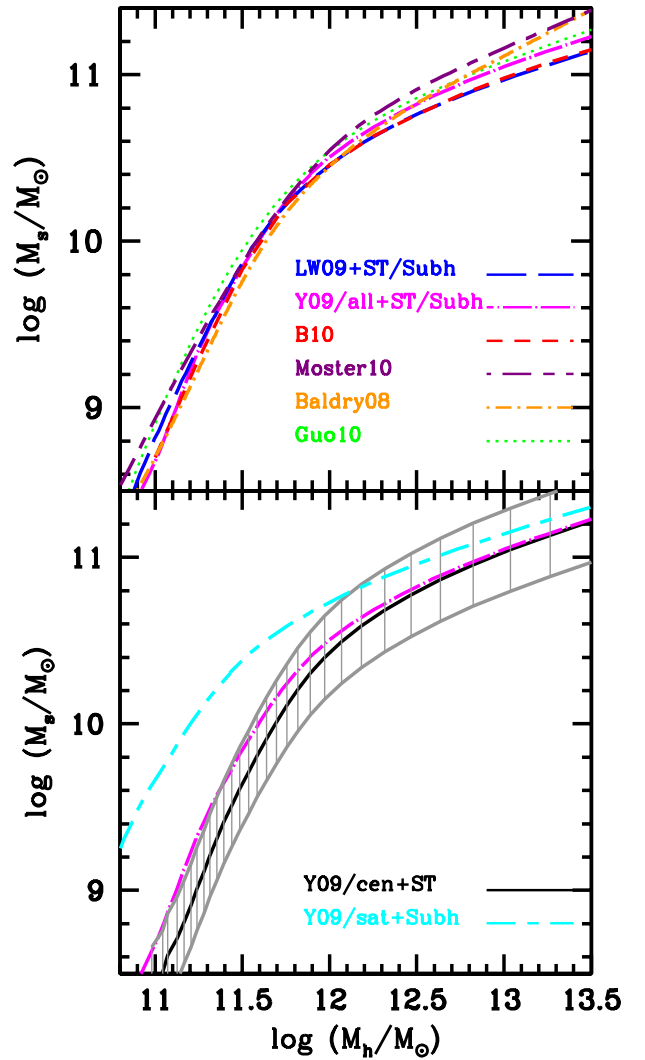}
\caption{\textit{Upper panel:} Stellar mass vs halo mass as inferred here by using
the Li \& White (2009) overall \gsmf\ and the S-T \HMF\ increased by the subhalo population
(long-dashed blue line) to be compared with the BCW10 inference, who used the same
\gsmf\ (short-dashed red line). The dot-dashed pink line shows the same \ms\ vs \mh\ inference
but by using the overall YMB09 \gsmf.  
Different determinations of the overall \ms--\mh\ relation by other authors (indicated in the
panel), who took into account in different ways the issue of group/cluster masses (see text) 
are also plotted \textit{Lower panel:} Same \ms--\mh\ relation as in the upper panel
(dot-dashed pink line) but for the central-only YMB09 \gsmf\ and the S-T (distinct) \HMF\ (solid line).
The grey curves connected by vertical lines show the estimated $1\sigma$ uncertainty
for the latter case. The \ms--\mh\ relation inferred for the only satellite YMB09 \gsmf\ and
the Giocoli et al. (2010) $z=0$ sub-halo mass function is plotted with the short-long-dashed
cyan line.
}
\label{MsMhrel}
\end{figure}

In Fig. \ref{MsMhrel}, the \ms--\mh\ relation obtained by using the Li \& White (2009) \gsmf\  
(see \S\S 2.1 and Fig. \ref{GSMFs}) and the S-T \HMF\ corrected to include sub-halos is plotted 
(long-dashed blue line). The relation given by BCW10, who also used as input the Li \& White (2009) \gsmf, 
is shown (short-dashed red line). Both curves are almost indistinguishable, showing an excellent
consistency of our results with those of BCW10 in spite of the differences in some of the 
methodological aspects.  

Further, we plot in Fig. \ref{MsMhrel} the \ms--\mh\ relation as above but using now
the total YMB09 \gsmf\ (dot-dashed pink line). This relation is similar to the one 
inferred using the Li \& White (2009) \gsmf. For $\log(\mh/\msun)\grtsim 12$, the 
former shifts with mass slightly to higher values of \ms\ for a given
\mh\ than the latter (at $\log(\mh/\msun)=13.5$ the difference is not larger than 0.08 dex in log\ms).
Such a shift is explained by the (small) systematical difference between the YMB09 and Li 
\& White {\gsmf}s at masses larger than $\log(\ms/\msun)\sim 11$ (see \S\S 2.1 and Fig. \ref{GSMFs}).

In Fig. \ref{MsMhrel}, the \ms--\mh\ relations given in Baldry et al. 
(2008, dot-dashed orange line), Moster et al. (2010, short-long-dashed line) and 
Guo et al. (2010, dotted green line) are also plotted.  When it was necessary, we have 
corrected the stellar masses to the Chabrier IMF, and the halo masses to the definition
of virial mass used here (see \S\S 2.2). As mentioned above, Baldry et al. corrected 
their \HMF\ to exclude groups/clusters of galaxies (something that we do but only
for the central blue  galaxies, see \S\S 2.2.1 and the result below).  As seen in 
Fig. \ref{MsMhrel}, their correction produces a steeper \ms--\mh\ relation at the 
high-mass side than in our case. Moster et al. and Guo et al. constrained the \ms--\mh\ 
relation by assigning stellar masses to the halos and subhalos of an $N$-body 
cosmological simulation in such a way that the total \gsmf\ is reproduced. 
Therefore, by construction, their \ms--\mh\ relations take into account the  
group/cluster halo masses issue. The \ms--\mh\ relations in both works are also 
slightly steeper than ours at high masses  but shallower on average than that 
one of Baldry et al. (2008).  Note that in BCW10 the scatter in \ms\ at fixed \mh\ 
was taken into account but the group/cluster halo masses issue was not. 

The \ms--\mh\ relation using the YMB09 \gsmf\ only for central galaxies and the distinct (S-T)
\HMF\ is plotted in the lower panel of Fig. \ref{MsMhrel} (solid black line). At large masses, 
this relation is quite similar to the one for all galaxies/satellites and halos/sub-halos (dot-dashed pink line). 
This is because at large masses the great majority of galaxies are centrals and the correction 
by sub-halos is negligible (see Figs. \ref{GSMFs} and \ref{HMFs}). 
At lower masses, the exclusion of satellites and sub-halos implies a lower \ms\ for a given
\mh. This is because the \gsmf\ decreases more than the \HMF\ as the mass is smaller when
passing from the total (galaxy and halo) samples to the central-only galaxy/distinct halo samples. 
The physical interpretation of this result could be that satellite galaxies of a given \ms\ have
less massive halos than central galaxies due to tidal stripping.  The \ms--\mh\ relation
derived only for the satellites YMB09 \gsmf\ and the Giocoli et al. (2010) $z=0$ sub-halo \HMF\ is
plotted in the lower panel of Fig. \ref{MsMhrel} (short-long-dashed cyan line).

\subsubsection{Uncertainties}

The uncertainty (standard deviation) in the \ms--\mh\ relation obtained by using 
the YMB09 central \gsmf\ and the distinct S-T \HMF\ (solid line),
is plotted in Fig. \ref{MsMhrel} (grey curves connected by vertical lines). As remarked in \S 2, 
we did not take into account all possible uncertainty sources in the \ms--\mh\ relation but have
just considered the two following ones:

 {\it (i)} The systematic uncertainty in stellar mass estimates, which 
is an uncertainty in the \gsmf. We assume  for this uncertainty a scatter of 0.25 dex 
(Gaussian distributed) independent of mass, and propagate it
to the \ms--\mh\ relation (by far it results the dominant 
source of error in the relation obtained with the AMT, see below and BCW10).

{\it (ii)} The intrinsic scatter in stellar mass at a fixed halo mass,
which is an uncertainty in the process of matching abundances.
To take into account this scatter in \ms\ at fixed \mh\ a probability
density distribution should be assumed. The convolution of this
distribution with the  true or intrinsic
\gsmf\ gives the  measured \gsmf. 
The cumulative true \gsmf\ is then the one used for the AM (BCW10).
The observational data allow to estimate the scatter in luminosity
(or \ms) and to date it appears to be independent of \mh\ {\bf (More et al.
2009; Y2009)}. In BCW10 a log-normal mass-independent scatter in 
\ms\ of 0.16$\pm0.04$ is assumed. Here, we follow the overall
procedure of BCW10 for taking into account this scatter.

We also explored the effect of {\it (iii)} the statistical uncertainty in the number density of the 
\gsmf\ (as given in YMB09), but we have found that the effect is negligible as compared to the 
one produced by item (i) (see also BCW10, their \S\S 4.3.1).  
The effect of the intrinsic scatter in \ms\ for a given \mh\ is also very small in the 
overall scatter of the \ms--\mh\ relation but it affects the high mass end of the
calculated \ms--\mh\ relation, where 
both the \gsmf\ and \HMF\ decay exponentially, since there are more low mass
galaxies that are scattered upward than high mass galaxies that are scattered 
downward (BCW10). For instance, at $\mh=10^{13.5}$ \msun, the stellar
mass after including this scatter is 1.2 times smaller. 
The contribution from all other sources of error, including uncertainties in the 
cosmological model, is much smaller ranging from 0.02 to 0.12 dex at $z=0$.

From Fig. \ref{MsMhrel} we see that the $1\sigma$ uncertainty in the \ms--\mh\ 
relation is approximately 0.25 dex in log\ms\ without any systematic dependence
on \mh, in good agreement with previous results (BCW10; Moster et al. 2010).
This uncertainty is larger 
than the differences between the \ms--\mh\ average relations found by different 
authors, including those that use the indirect AMT but with different {\gsmf}s, 
methodologies, and corrections, and those who use more sophisticated formalisms
(see for comparisons and discussions BCW10 and More et al. 2011). 
On one hand, this shows that most methods and recent studies aimed at 
relating halo masses to observed galaxies as a function of their stellar masses
are converging to a relatively robust determination. On the other hand,  this result 
suggests that attaining a higher precision in estimating \ms\ from observations is 
the crucial task for lowering the uncertainty in the inference of the \ms--\mh\ relation.

\begin{figure}
\vspace{11.5cm}
\includegraphics{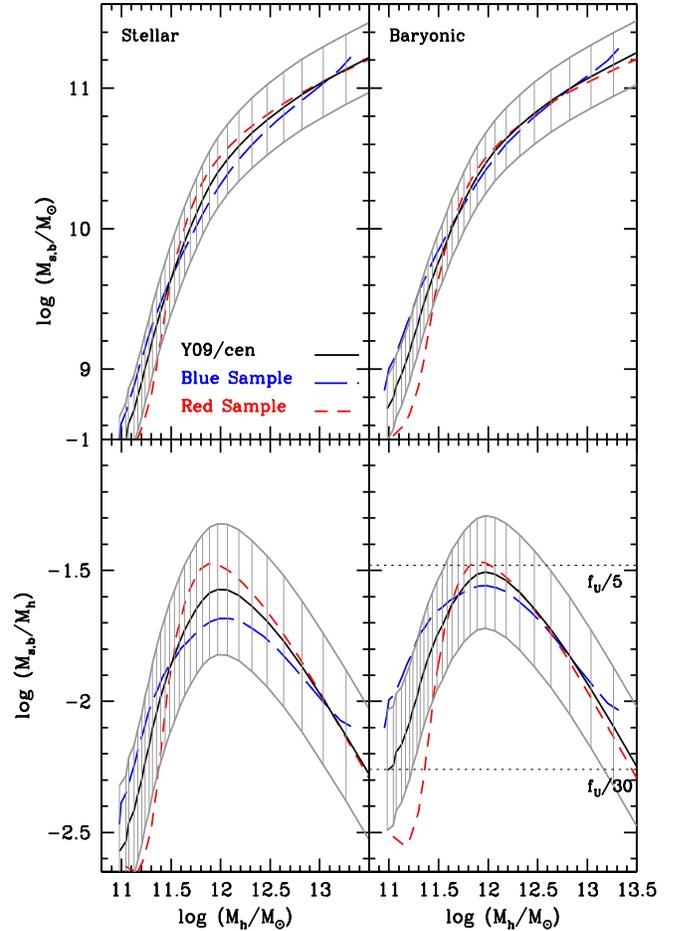}
\caption{\textit{Left panels:} Mean \ms--\mh\ (top) and \fs--\mh\ (down) relations of all central
(solid black line), blue central  (long-dashed blue line), and red central (short-dashed red line) galaxies
as inferred here by using the YMB09 data. The grey curves connected by vertical lines show the 
1$\sigma$ uncertainty for the all galaxies case; similar uncertainty regions around the main relations
are found for the blue and red sub-samples (see Fig. \ref{comp}). \textit{Right panels:} Same as in 
left panels but for \mb\ instead of \ms. Dotted lines: $\fb=f_U/5$ and 
$f_U/30$, where $f_U=0.167$ is the universal baryon fraction. 
}
\label{types}
\end{figure}


\subsection{The stellar-halo mass relations for central blue and red galaxies}

The upper and lower left panels of Figure \ref{types} show the mean \ms--\mh\ 
and \fs--\mh\ relations for: all central galaxies (solid line, as in Fig. \ref{MsMhrel}), 
central blue (short-dashed line), and central red (long dashed line) galaxies. 
In order to infer these relations for blue galaxies, the central blue YMB09 {\gsmf} and 
the distinct (S-T) \HMF\ corrected for excluding halos (i) associated to observed 
groups/clusters of galaxies and (ii) that suffered central major mergers since $z=0.8$ 
(see \S\S 2.2.2) 
were used. In the case of red galaxies, the central red YMB09 {\gsmf} and the \HMF\ 
complementary to the one associated to blue galaxies were used.  

The shaded area in Figure \ref{types} is the same $1\sigma$ uncertainty 
showed in Fig. \ref{MsMhrel} for the overall central sample. The uncertainties 
corresponding to the \ms--\mh\ and \fs--\mh\ relations for the blue and red 
galaxy sub-samples would be close to the one of the total sample in case 
the corrections made to the \HMF\ do not introduce an extra uncertainty.
In fact this is not true, in particular for the group/cluster mass function
introduced to correct the \HMF\ associated to blue galaxies. Unfortunately,
the work used for this correction does not report uncertainties. Hence,
the uncertainties calculated here for the blue and red samples (shown
explicitly in Fig. \ref{comp} below) could be underestimated, specially at 
large masses. 


\begin{figure}
\vspace{11.5cm}
\includegraphics{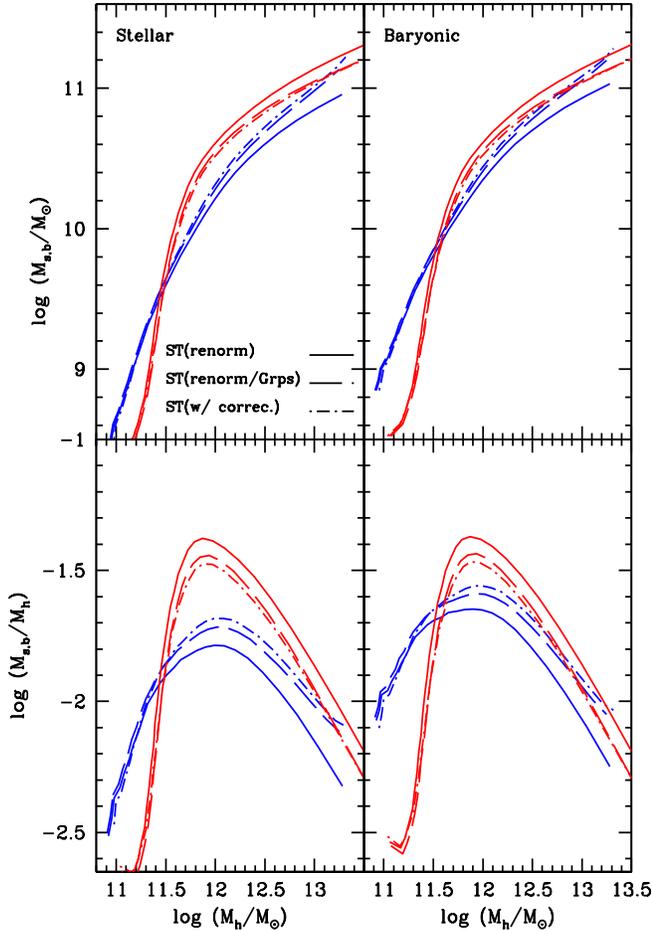}
\caption{\textit{Left panels:} Mean \ms--\mh\ (top) and \fs--\mh\ (down) relations of central blue 
(blue lines) and red (red lines) galaxies  when (i) no systematical corrections to the corresponding 
"blue" and "red" {\HMF}s were applied apart from re-normalisations in the global abundance 
(see text, solid lines), (ii) the {\HMF}s were corrected by group/cluster abundances and 
re-normalised  (long-dashed lines), and (iii) the {\HMF}s were corrected both
by group/cluster abundances and late major mergers (as in Fig. \ref{MsMhrel}, 
dot-dashed lines). \textit{Right panels:} Same as in left panels but for \mb\ instead of \ms. 
}
\label{types2}
\end{figure}


In the mass range $11.5\lesssim$log($\mh/\msun$)$\lesssim 13.0$, 
the \ms--\mh\ and \fs--\mh\ relations for central blue (red) galaxies lie slightly
below (above) the relations corresponding to the overall sample. For masses below these
ranges, the trends invert. The \fs--\mh\ curves for blue and red sub-samples peak at 
log($\mh/\msun$)$=11.98$ and 11.87, with values of $\fs=0.021^{+0.016}_{-0.009}$ 
and $\fs=0.034^{+0.026}_{-0.015} $, respectively.  The corresponding stellar 
masses at these peaks are log($\ms/\msun$)$=10.30\pm0.25$ for blue galaxies and 
log($\ms/\msun$)$= 10.40\pm0.25$ for red galaxies. These masses are around 
0.23 and 0.30 the characteristic stellar mass $M^\star\approx 10^{10.93}\msun$ of the 
overall YMB09 \gsmf, respectively. The maximum difference between the blue and red mean \ms--\mh\ 
relations is attained at log($\mh/\msun$)$\approx 11.9$; at this mass, the \fs\ value of the 
former is 1.7 times smaller than the \fs\ of the latter. For larger masses this difference 
decreases. 

At the low-mass end,  roughly $\fs\propto M_h$ ($\propto M_s^{0.5}$) and 
$\fs\propto M_h^{3.0}$ ($\propto M_s^{0.8}$) for the blue and red samples, 
respectively, while at the high-mass end, $\fs\propto M_h^{-0.4}$ ($\propto M_s^{-0.7}$) 
and $\fs\propto M_h^{-0.6} $ ($\propto M_s^{-1.5}$), respectively.

It is important to note that the differences between blue and red \ms--\mh\ relations 
at almost all masses are within the $1\sigma$ uncertainty of our inferences. We 
conclude that the \ms--\mh\ (\fs--\mh) relation does not depend significantly on galaxy 
color (type).  If any, the mean \fs--\mh\ relation of red galaxies is narrower and more
peaked than the one of blue galaxies. 
In the mass range where the abundances of blue and red galaxies are closer 
($10.0<\log$(\ms/\msun)$<10.7$), the intrinsic scatter around 
the \ms--\mh\ relation would slightly correlate with color in the sense that the redder 
(bluer) the galaxy, the larger (smaller) is its \ms\ for a fixed \mh, with a maximum 
average deviation from the mean due to color not larger than $\sim 0.1$ dex.
For masses smaller than $\ms\approx 10^{9.7}$ \msun, the correlation of the
scatter with color would invert. 

The (slight) differences between blue and red \ms--\mh\ (\fs--\mh) relations can be understood
basically by the differences in the respective cumulative {\gsmf}s and, at a minor level, by the 
differences of the corresponding {\HMF}s for each case. The sharp peak in the red 
\fs--\mh\ relation is associated to the turn-over at $\ms\sim10^{10.5}$ \msun\ in the 
\gsmf\ of red galaxies (see Fig. \ref{GSMFs}). 

In order to estimate the influence of the corrections introduced to the \HMF\ for blue 
(red) galaxies,  we have redone the analysis by using the original distinct (S-T) \HMF\ 
without any correction but re-normalised to obtain the same fraction of halos as the 
fraction implied by the \gsmf\ of blue (red) galaxies with respect to the total \gsmf.
The results are shown in Fig. \ref{types2} with solid curves of blue color 
(blue galaxies) and red color (red galaxies). 
For comparison, the corresponding relations plotted in Fig. \ref{types}
are reproduced here (dot-dashed blue and red lines, respectively). One sees
that the corrections to the \HMF\ we have introduced for associating
halos to the blue and red galaxy sub-samples act in the direction of
reducing the differences among them in the \ms--\mh\ (\fs--\mh) relations,
specially for larger masses.  The group/cluster mass function correction
to the \HMF\ hosting central blue galaxies is the dominant one.
The dashed blue and red curves show such a case, when only this
correction (and a small re-normalisation) is applied.

\begin{table}
\caption{Fit parameters}
\begin{center}
\begin{tabular}{|c|c|clc}
\hline
 Parameter &  All & Blue & Red \\
\hline
 $\log M_{0,h}$ & 11.97& 11.99 &11.87 \\ 
 $\log M_s^*$ & 10.40 & 10.30 & 10.40 \\ 
 $\beta$            &   0.34 &  0.37  & 0.18 \\   
 $\alpha$          &   1.45 & 0.90 & 1.50 \\
$\gamma$        &  0.90  & 0.90 & 0.90 \\
$a$ ($M_s<M_s^*$)                 &  0.000       & 0.125 & 0.000  \\
$a$ ($M_s>M_s^*$)                &  0.095 &  0.125 & 0.093 \\
\end{tabular}
\end{center}
\end{table}%

\subsubsection{Analytical fits to the stellar--halo mass relations}

From the comparison of the \gsmf\ and \HMF\ it is easy to deduce that high- and low-mass 
galaxies have significantly different  \ms--\mh\ scalings, a fact attributed to the different
feedback/gas accretion mechanisms dominating in large and small systems (see e.g.,
Benson et al. 2003). The transition point between the low- and high-mass scalings
defines a characteristic halo mass $M_{0,h}$ and an associated stellar mass $M_s^*$.
Therefore, it was common to describe the \ms--\mh\ relation as a double-power law with 
the turnover point at $M_{0,h}$. However, BCW10 have argued recently that a power-law
at the high-mass side is conceptually a bad description for the \ms--\mh\ relation and 
proposed a modification to it. Our results show indeed that a power-law is not enough to
describe the high-mass side of the \ms--\mh\ relations.

\begin{figure}
\vspace{7.7cm}
\includegraphics{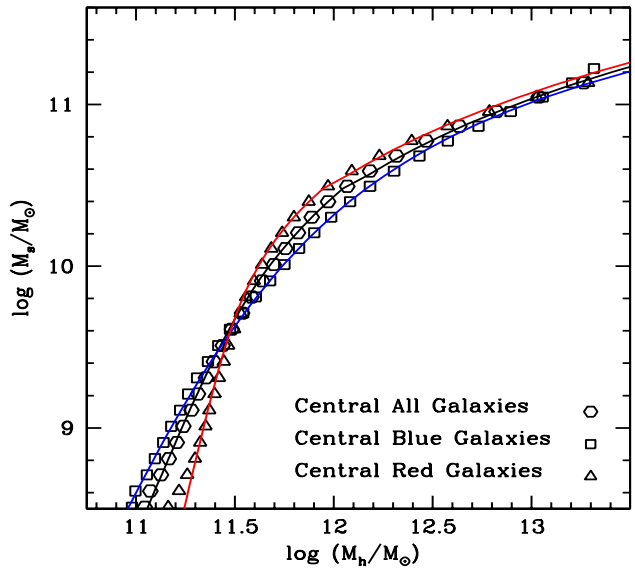}
\caption{Analytical fits given by eq. (\ref{fitMhMs}) and Table 1 compared to
the mean \ms--\mh\ relation obtained here for all central galaxies (black
solid line) and the central blue (blue solid line) and red (red solid line)
galaxy sub-samples.
}
\label{fits}
\end{figure}


We have found that a good analytical description to the overall, blue, and red mean \ms--\mh\ relations
inferred here can be obtained for the inverse of the relations (\mh\ as a function of \ms, as in BCW10) 
by proposing a power-law dependence for low masses and a sub-exponential law for high masses
(see BCW10). The functional form that fits well the three \mh--\ms\ relations is:   
\begin{equation}
M_h=\frac{M_{0,h}}{2^\gamma} \left[ \left(\frac{M_s}{M_s^*}\right)^{\beta/\gamma}+\left(\frac{M_s}{M_s^*}\right)^{\alpha/\gamma} \right]^\gamma10^{a(M_s/M_s^*-1)}
\label{fitMhMs}
\end{equation}
where $\beta$ regulates the behaviour of the relation at masses $\ms<M_s^* $, $\alpha$ together with  
the sub-exponential term ($a<1$) regulate the behaviour at masses $\ms>M_s^*$, and $\gamma$ 
regulates the transition of the relation around $M_s^*$. In Table 1 are given the values of all the
parameters that best fit our results for the (central) overall, blue, and red \ms-\mh\ relations. Note that $a$ 
assumes two different values depending on whether the  mass is smaller or larger than $M_s^*$.

Figure \ref{fits} shows the three mean \ms--\mh\ relations obtained here and the functional
form given in eq. (\ref{fitMhMs}) with the corresponding parameters reported in Table 1. 
The functional form is an excellent fit to the overall and blue \ms--\mh\ relations at all masses
and to the red  \ms--\mh\ relation for masses larger than $\mh\approx 10^{11.3}$ \msun.

\subsection{The baryonic-halo mass relations for central blue and red galaxies}

The right upper and lower panels of Fig. \ref{types} show the mean \mb--\mh\ and 
\fb--\mh\ relations, as in the left panels, for all central galaxies (solid line), 
central blue (long-dashed blue line), and central red (short-dashed red line) 
galaxies. The blue and red {\gbmf}s were calculated from  the corresponding {\gsmf}s
and adding to \ms\ the respective gas mass, \mg\ (see \S\S 2.1). 
The total {\gbmf} is the sum of both of them.
The error in \mb\ was calculated as the sum 
in quadratures of the errors in \ms\ and \mg. This error, together with the intrinsic scatter in 
\ms\ (see \S\S 2.2), both propagated to the \mb--\mh\ relation, account for an 
uncertainty (standard deviation) of $\sim 0.23$ dex in log\mb\ at all masses (grey
curves connected by vertical lines in Fig. \ref{types}). 

The baryonic mass fraction, \fb, for blue galaxies is larger than the corresponding 
stellar one, \fs, in particular at smaller halo masses. At $\mh\approx 10^{11}$ \msun, 
\fb\ is a factor 2.4 times higher than \fs, while the peak of $\fb= 0.028^{+0.018}_{-0.011}$
(at $\mh= 10^{12.0}$ \msun) is only 1.3 times larger than the peak of \fs\ 
(at $\mh= 10^{12.0}$ \msun). For larger masses, the difference between 
\fb\ and \fs\ decreases, while for smaller masses, the lower is \mh, the
larger is \fb\ than \fs. For red galaxies, \fs\ and \fb\ are very similar, some differences
being observed only at the lowest masses.

For masses larger (smaller) than $\mh\approx 10^{11.6}$ \msun, the differences  between 
the \mb--\mh\ (\fb--\mh) relation of blue and red galaxies become smaller (larger) than 
in the case of stellar masses (see \S\S 3.2 and left panels of Fig. \ref{types}). In general,
the \fb\ bell-shaped curve for red galaxies is more peaked and narrower than the one
for blue galaxies.

For blue galaxies, roughly $\fb\propto \mh^{0.7}$ ($\mb^{0.4}$) at the low-mass end, and  
$\fb\propto \mh^{-0.5}$ ($\mb^{-0.8}$) at the high-mass end.  For red galaxies,  roughly 
$\fb\propto \mh^{2.9}$ ($\mb^{0.8}$) at the low-mass end, and $\fb\propto \mh^{-0.6}$ ($\mb^{-1.5}$)
 at the high-mass end. For halos of masses 
$\mh\approx  10^{11.0}$ \msun\ and $\mh\approx 10^{13.2}$ \msun, the baryon 
fraction for blue (red) galaxies decreases to values $\fb\approx 0.004$ and 0.0085 
($\fb\approx 0.0031$ and 0.0071),  respectively. 
Therefore, for all masses, $\fb<<f_U$, where $f_U\equiv\Omega_b/\Omega_M$ is the universal 
baryon mass fraction; for the cosmology used here, $f_U=0.167$.

\section{Discussion}

\subsection{Comparison with other works}

As discussed in \S 3.1 (see Fig. \ref{MsMhrel}), our inference of the local overall
\ms--\mh\ relation is in general in good agreement with several recent works 
that make use of the AMT (e.g., Baldry et al. 2008; Guo et al. 2010; 
Moster et al. 2010; BCW10). 
The aim in this paper was to estimate the \ms--\mh\ and \mb--\mh\
relations for blue (late-type) and red (early-type) central galaxies
separately. We have found that the differences between the means
of the obtained relations for blue and red galaxies are within the 
1$\sigma$ uncertainty (see Fig. \ref{types}). In more detail, the mean stellar and 
baryonic mass fractions (\fs\ and \fb) as a function of \mh\ for red galaxies are 
narrower and more peaked than those for blue galaxies in such a way that in a given mass range
(11.5--13.0 and 11.5--12.5 in log($\mh/\msun$) for the stellar and baryonic
cases, respectively) the former are higher than the latter and outside these ranges, 
the trend is inverted, specially at the low--mass side. 

There are only a few previous attempts to infer the halo masses of central 
galaxies as a function of mass (luminosity) \textit{and} galaxy type (Mandelbaum 
et al. 2006; More et al. 2011). These works use direct techniques (see Introduction), 
which are, however, limited by low signal-to-noise ratios, specially for less massive 
systems in such a way that the halo mass determinations are reliable only
for galaxies with $\ms\grtsim 10^{10}$ \msun. These techniques are 
galaxy-galaxy weak lensing and kinematics of satellite galaxies around central 
galaxies. In order to overcome the issue of low signal-to-noise ratios in the 
current measures, large samples of galaxies are stacked together in bins of 
similar properties (e.g., luminosity, \ms, galaxy type) obtaining this way 
higher (statistically averaged) signals of the corresponding measures 
(the tangential shear in the case of lensing and the weighted satellite velocity 
dispersion in the case of satellite kinematics). Besides, estimates of \mh\ 
with these sophisticated techniques are subject to several assumptions, among them, 
those related to the internal halo mass distribution. It is usual to assume 
the Navarro, Frenk \& White (1997) density profile with the mean concentration 
for a given mass as measured in N-body cosmological simulations.

It is not easy to achieve a fair comparison of the results obtained with the AM 
formalism and those with the direct methods. We have inferred the mean (and scatter) 
of log\ms\ as a function of \mh, while the weak lensing and satellite kinematics 
techniques constrain \mh\ as a function of \ms\ (see e.g., More et al. 2011); besides,
the former calculates the mean of \mh\ (and its scatter) in a linear scale instead of
a logarithmic one. These different ways of defining the relationship 
between stellar and halo masses, depending on the shapes and scatters of the 
corresponding relations, diverge less or more among them. In BCW10
(see their Fig. 10), it was shown that at low masses ($\log(\mh/\msun\lesssim 12$,
$\log(\ms/\msun)\lesssim 10.5$), averaging log\ms\ as a function
of \mh\ or log\mh\ as a function of \ms\ give equivalent results 
for the AMT, but at high masses, where the \ms--\mh\ relation becomes 
much shallower, this relation becomes steeper (higher stellar mass at a fixed 
halo mass) for the latter case with respect to the former one.

\begin{figure*}
\vspace{215pt}
\includegraphics{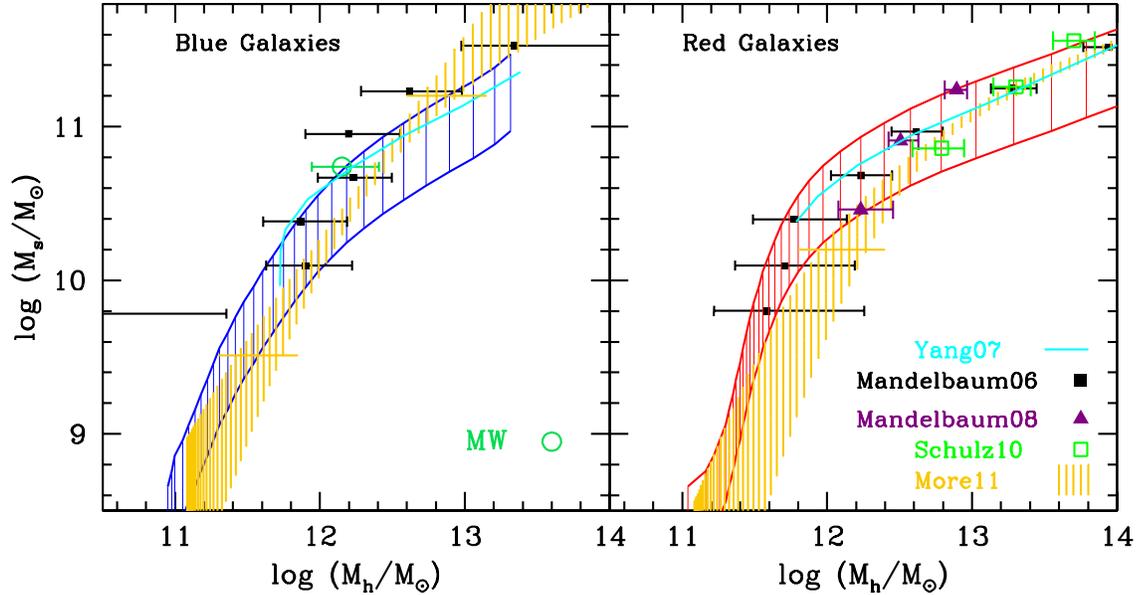}
\caption{Comparison with other observational inferences. \emph{Left panel}: 
$\ms-\mh$ relation for blue (late-type) galaxies. The blue curves connected 
by vertical lines encompass the $\pm 1\sigma$ interval inferred here. We also
reproduce the inferences using galaxy-galaxy weak lensing by Mandelbaum et al. (2006, 
black squares),  galaxy groups (Yang et al 2007, cyan solid line), and satellite kinematics 
(More et al 2011, orange vertical lines). Estimates for the Milky Way are plotted
(open circle with error bar). \emph{Right panel}:  $\ms-\mh$ relation for red (early-type) 
galaxies. The red curves connected by vertical lines encompass the $\pm 1\sigma$ interval 
inferred here. Other determinations as in the left panel but for early-type
galaxies are plotted. More recent inferences with the weak lensing technique by Mandelbaum et al.  
(2008, filled violet triangles) and by Schulz et al. (2010, open green squares) are also plotted.
}
\label{comp}
\end{figure*}

In Fig. \ref{comp}, the results from Mandelbaum et al. (2006) are reproduced, 
left panels for central late-type galaxies and right panels for central 
early-type galaxies (solid squares with error bars). The error bars are 95 
percent confidence intervals (statistical). Mandelbaum et al. (2006) 
have used the (de Vaucouleours/exponential) bulge-to-total ratio, frac\_deV, 
given in the SDSS PHOTO pipeline as a criterion for late- ({frac\_deV}$<0.5$) 
and early-type ({frac\_dV}$\ge 0.5$) separation. This criterion of course 
is not the same as the color used in YMB09, but there is a correlation between 
both of them in such a way that a comparison between our results and those of Mandelbaum 
et al. is possible at a qualitative level. Note that we have diminished the 
halo masses of Mandelbaum et al. by $\approx 15\%$ on going from their 
to our definition of halo virial mass.  In more recent works, Mandelbaum et al. 
(2008) and Schulz et al. (2010) reported a new weak lensing analysis for the 
massive central early-type galaxies using the seventh SDSS data release (DR7)
and a more sophisticated criteria for selecting the 
early-type lens population. Their results are plotted in the right panel of 
Fig. \ref{comp} with solid triangles and open squares, respectively.

In the case of the satellite kinematics determinations of \mh\ by More et al. 
(2011), the same SDSS sample and similar recipes as in YMB09 for calculating 
\ms, classifying galaxies into blue and red, and finding central and satellites 
galaxies were used.  More et al. (2011) applied their analysis to constrain
the mean log\mh\ as a function of \ms, but also present the constraints of their 
model for the mean of log\ms\ as a function of \mh. Their results for the latter case, 
kindly made available to us in electronic form by Dr. S. More, are reproduced in 
Fig. \ref{comp} as the shaded (orange) regions which represent the 68\% 
confidence intervals. On going from their to our definitions of halo mass 
and IMF, their \mh\ and \ms\ were diminished by $\approx 15\%$ and  
$\approx 25\%$, respectively. The dotted horizontal lines in each panel
show the approximate range in \ms, where the determinations are
reliable according to More et al. (2011; see their Fig. 11).

In More et al (2011) are also reported results for the average \mh\ as a 
function of \ms\ split in central blue and red galaxies corresponding to 
the galaxy group analysis by Yang et al. (2007). The solid (cyan) curves 
in Fig. \ref{comp} reproduce these results.

Finally, the standard $\pm 1\sigma$ deviation intervals that we have obtained 
from the AMT are reproduced in Fig. \ref{comp} for central blue and
red galaxies (solid blue and red curves connected by vertical lines, respectively). 
Note that in the determinations with direct methods, the systematic uncertainty in \ms, 
which is the main source of error in the AMT, was not taken 
into account. 

Our inference for early-type (red) galaxies is consistent (within the uncertainties, 
errors, and different ways of presenting the constraints) with the weak 
lensing results of Mandelbaum et al. (2006) and Schulz et al. (2010), and 
with the galaxy group analysis of Yang et al. (2007) as reported in More et 
al. (2011), for all the masses reported in each one of these papers. With 
respect to the satellite kinematics analysis by More et al. (2011), their mean 
halo masses for $\ms\sim 5\times 10^{9}-10^{11}$ \msun\ 
(for smaller masses their uncertainties are very large), are larger than ours
(and those of Mandelbaum et al. 2006) by factors around $2$. 
For larger masses, all determinations agree roughly with our results. 
In fact, there is some indication that satellite kinematics yields halo 
masses around low mass central galaxies that are systematically larger than most 
other methods, specially for red central galaxies (Skibba et al. 2011; but 
see More et al. 2011 for a discussion). 

For late-type (blue) galaxies, our results are in reasonable agreement with 
those of  Mandelbaum et al. (2006) for masses $\ms\lesssim 10^{10.8}$ \msun.  
At higher masses, their results imply halo masses for a given \ms\ smaller than ours,
with the difference increasing to higher stellar mass.  
The discrepancy would be weaker if taking into account that the mean \ms--\mh\ 
relation in our case becomes steeper when calculating \mh\ as a function of 
\ms. On the other hand, it must be said that the 
number statistics becomes poor for massive late-type galaxies, resulting in a 
stacked weak lensing analysis with large error bars. For example, in the two 
most massive bins in the Mandelbaum et al. (2006) sample (the two uppermost 
points in Fig. \ref{comp}), only 5 and 11 percent of the galaxies are classified 
as late types.  Future weak lensing works should confirm whether high-mass 
late-type galaxies have or not such relatively small halos as found in Mandelbaum
et al. (2006).  Regarding the comparison with the satellite kinematics inferences 
of More et al. (2011), the agreement is reasonable at least up to $\ms\approx 10^{11}$ 
\msun, though the relation inferred by these authors is less curved than ours.
For larger masses, these authors caution that their results become 
very uncertain, as in the weak lensing case, because of poor statistics of 
massive blue galaxies. The galaxy groups (Yang et al. 2007) inference,
in the mass range allowed by this technique, gives halo masses slightly smaller 
than the means of our inference for a given \ms.

In general, most techniques for inferring the 
relationship between stellar and halo masses of galaxies agree among them
within factors up to 2--3 in \mh\  (BCW10; More et al. 2011; 
Dutton et al. 2010).  This seems to be also the case for samples 
partitioned into late- and early-type galaxies, as shown here. However, beyond 
the detailed comparison between our results and those obtained with direct 
techniques, it seems that there is a systematic qualitative difference:
in our case, at a given halo mass (for $10^{11.5}$ \msun $\lesssim \mh \lesssim 10^{13.0}$ \msun),  
blue centrals, on average, have lower stellar masses than red centrals, 
while in the case of determinations with direct techniques, the opposite 
applies at least for masses larger than $\mh\sim 10^{12}$ \msun\ 
(Mandelbaum et al. 2006; More et al. 2011; see also Figs. \ref{types} and \ref{comp}).

A partial source of bias contributing to this difference could be that in the 
weak lensing and satellite kinematics techniques the same concentration for halos 
hosting late- and early-type galaxies is assumed. If halos of late- (early-)type 
galaxies are less (more) concentrated than the corresponding average, then for 
the same measure (shear or satellite velocity dispersion), the halo masses are 
expected to be higher (lower) than the obtained ones. Therefore, the differences 
found (Mandelbaum et al. 2006 and More et al. 2011) in the mass halos of 
late- and early-type galaxies of a given \ms\ would decrease or even invert their sense.  

While it is difficult to make any robust statement about possible systematics 
in one or another technique regarding late and early types, we ask ourselves 
what should be modified in our assumptions in order to invert the behaviour 
of the \ms--\mh\ relations with galaxy type (color) obtained here. We have shown
in Fig. \ref{types2}  that our corrections to the \HMF\ had the effect of making closer
the \ms--\mh\ relations of blue and red galaxies at large masses. One possibility
in order not only close more the relations but invert them
is to make even steeper (shallower) the \HMF\ corresponding to blue (red) 
galaxies, mainly at the high-mass end (see Fig. \ref{HMFs}, lower panel). 
This would imply, for instance, a higher correction to the \HMF\ due to groups
than that made by us. The group/cluster mass function used by
us (Hein{\"a}m{\"a}ki et al. 2003) is one of the most general ones found in 
the literature; it includes all kinds of groups/clusters with 3 or more members 
and $\delta N/N\ge 80$. The authors note that their sample is complete
down to a dynamical mass roughly equivalent to $\mh=5\times 10^{13}$ \msun. 
It could be that the abundance of groups of lower masses is larger
than that given in Hein{\"a}m{\"a}ki et al. (2003), though it is difficult to
accept that blue galaxies are completely absent in the centers of
small and loose groups of a few ($>2$) members. 

Last but not least, in Fig. \ref{comp} we include observational estimates for our Galaxy (open circle). 
The uncertainties in the estimates of \mh\ for the Milky Way are still large but 
better than most of the determinations for other individual galaxies. For recent 
reviews on different results see Guo et al. (2010) and 
Dutton et al. (2010). In Fig. \ref{comp} we plot a recent estimate of \mh\ based on 
observations of 16 high velocity stars (Smith et al 2007). These authors find 
$\mh=1.42^{+1.14}_{-0.54}\times10^{12}\msun$, which is in good agreement with 
several previous works (e.g., Wilkinson \& Evans 1999; Sakamoto et al.
2003; Li \& White 2008), though results from Xue et al. (2008) suggest lower values (but see a
recent revision by Przybilla et al. 2010).  For its \mh, the \ms\ of Milky Way
seems to be in the high extremum of blue galaxies, close to values typical of red 
galaxies. It should be said that it is an open question whether the Milky Way
is an average galaxy or not. In the stellar Tully-Fisher and radius--\ms\ relations 
(e.g., Avila-Reese et al. 2008), the Milky Way is shifted from the 
average to the high-velocity and low-radius sides, respectively.

\subsection{Interpretations and consistency of the results}

Although our main result is that the differences between the \ms--\mh\ and \mb--\mh\ relations
for central blue and red galaxies are marginal (within the uncertainties of our determinations), 
we will explore whether such differences are expected or not.  For this it is important to 
approach the problem from an evolutionary point of view. 

In Firmani \& Avila-Reese (2010,
hereafter FA10), the determinations of the \ms--\mh\ relation for all galaxies at different
redshifts, out to $z=4$ (BCW10), and the average $\Lambda$CDM individual halo mass 
aggregation histories (MAHs) were used to determine the individual {\it average} 
\ms\ growth of galaxies in general as a function of mass (called in that paper as Galaxian
Hybrid Evolutionary Tracks, GHETs).  It was found that the more massive the galaxies,
the earlier transit from their active (star-forming, blue) regime of \ms\ growth to a passive (red) 
phase (population 'downsizing'), while their corresponding halos continue growing, more 
efficiently at later epochs as more massive they are ('upsizing'). 
The inferred trend for the transition stellar mass is log($\mtran/\msun$)$\approx 10.30 + 0.55z$.
Therefore, galaxies of mass $\ms\approx 10^{10.3}$ \msun\ are on average becoming passive
(red) today. For $\ms\grtsim \mtran$, the larger the mass, the redder will be the galaxy on average.
The opposite applies for  $\ms\lesssim \mtran$, the smaller the mass, the bluer will be the galaxy. 
Interesting enough, $\ms\approx 10^{10.3}$ \msun\ is roughly the mass where the overall 
YMB09 blue and red {\gsmf}s cross: for masses larger than this crossing mass, \mcross, 
redder galaxies become more and more abundant than bluer ones and the inverse 
happens at smaller masses (see Fig. \ref{GSMFs}). 

Galaxies that are transiting from active to passive at $z\sim 0$ (those around 
$\mtran\approx 10^{10.3}$ \msun) have probably been subject recently to a process 
that induced an {\it efficient} transformation of the available gas into stars in such a way 
that their stellar populations started to redden passively. Hence, for a given \mh, 
they are expected to have a higher \ms\ (or \fs) than those galaxies of similar 
mass that did not suffer (yet?) the mentioned above process (bluer ones). 
The relatively small difference in \fs\ for blue and red galaxies we have found 
here (whose maximum is attained namely around $\mtran\sim\mcross$, 
Fig. \ref{types}) would imply that the scatter around  \mtran\ is moderate. 

Galaxies more massive than \mtran\ (or \mcross), according to the evolutionary analysis 
by FA10, had the process of efficient gas consumption into stars (and the further cessation 
of \ms\ growth) earlier on average as more massive is the galaxy, while their halos continue 
growing. Therefore, one expects that the more massive the galaxy, the redder and the 
lower its stellar (and baryonic) mass fraction \fs\ will be on average. The few blue massive galaxies 
may have slightly smaller stellar masses (lower \fs) than the corresponding red ones 
because they should have transformed gas into stars less efficiently in the past. Therefore,
by including gas, i.e. when passing to \fb\,  the difference between blue 
and red massive galaxies at large masses should become negligible. This is
what indeed happens (see Fig. \ref{types}).

Galaxies less massive than \mtran\ (or \mcross) at $z\sim 0$, according to FA10, are 
in general more actively assembling their stellar masses as smaller they are ('downsizing
in specific SFR), while their dark halo mass growth is already very slow. 
This implies the existence of relatively larger reservoirs of cold gas in the galaxies 
as smaller they are (gas not related to the halo-driven infall) 
because the SF has been delayed in the disk and/or cold gas is being lately 
(re)accreted into the galaxy. However, if for some reason the gas reservoir in these 
galaxies is lost, then the galaxy will redden and its baryonic and stellar 
mass fractions will be smaller than of the galaxies that were able to keep their 
gas reservoir (the majority), in agreement with our inferences here (Fig. \ref{types}).

\section{Summary and Conclusions}

By means of the AM technique and using the central blue and red {\gsmf}s, 
constructed from the local SDSS sample by YMB09, we have inferred the local \ms--\mh\ (or \fs--\mh) 
relations for {\it central} galaxies and for the sub-samples of blue and red galaxies. 
To derive the relations for the sample of blue galaxies, (i)
the mass function of observed groups/clusters of galaxies is subtracted 
from the distinct (S-T) \HMF\ (blue, late-type galaxies are not observed in the 
centers of groups and clusters), and (ii) 
halos that suffered a major merger since $z=0.8$ are excluded.  
For red galaxies, the \HMF\ 
is assumed to be the complement of the ``blue" one, with respect to the 
overall (distinct) \HMF. 
We consider as sources of uncertainty in our analysis 
only the systematical error in assigning stellar masses to galaxies (0.25 dex)  and the 
intrinsic statistical scatter in stellar mass at a fixed halo mass (0.16 dex). By using the 
observational \mg--\ms\ relation and its scatter, we transited from \ms\ to \mb\ (=\ms + \mg) 
in the \gsmf\ and estimated the overall blue and red {\gbmf}s, which are used to obtain  
the corresponding baryonic \mb--\mh\ (or \fb--\mh) relations using the AM technique. 

The \ms--\mh\ relation obtained here agrees rather well with previous studies 
(see Fig. \ref{MsMhrel}). The small differences found in this work
can be explained mainly in terms of the different {\gsmf}s used in each study, and to a less extent by 
variations in the methodology. The $1\sigma$ uncertainty in the obtained 
\ms--\mh\ relation is $\approx 0.25$ dex in log\ms.  The  \ms--\mh\ relation
of central galaxies lies below (lower \ms\ for a given \mh) the
overall one by a factor $\sim 1.6$ at $\mh=10^{11}$ \msun\ and less than
5\% for $\mh>10^{13}$ \msun. 

Our main result refers to the calculation of the {\it central} \ms--\mh\ and \mb--\mh\ 
relations for the two broad populations into which the galaxy sample can be divided: blue (late-type) 
and red (early-type) galaxies. We highlight the following results from our analysis:

$\bullet$ 
At $\mh\grtsim 10^{11.3}$ \msun\ the mean stellar mass fraction \fs\ of 
blue galaxies is lower than the one of red galaxies, the 
maximum difference being attained at \mh$\approx 10^{11.7}$ \msun; at this 
mass, the \fs\ of red galaxies is 1.7 times the one of blue galaxies (see Fig. \ref{types}). 
At larger masses, the difference decreases until it disappears. At $\mh\lesssim 
10^{11.3}$ \msun\ the trend is reversed as blue galaxies tend to have higher values of \fs\ 
than red ones. In the case of the baryonic mass fractions, 
\fb, the same trends of the stellar relations remain but at $\mh\grtsim 10^{11.3}$ 
\msun\ the difference in \fb\ between blue and red galaxies is small, 
while for smaller masses, the difference increases. 

$\bullet$ 
The \ms--\mh\ and \mb--\mh\ (or \fs--\mh\ and \fb--\mh) relations of
central blue and red sub-samples do not differ significantly from the 
respective relations of the overall central sample, and these 
differences are within the $1\sigma$ uncertainty of the inferences 
(Fig. \ref{types}). For blue (red) galaxies, the maximum value of 
\fs\ is $0.021^{+0.016}_{-0.009}$ ($0.034 ^{+0.026}_{-0.015}$) 
and is attained in halos of mass 
$\mh= 10^{11.98}$ \msun\ ($\mh= 10^{11.87}$ \msun); the corresponding 
stellar mass is $\ms= 10^{10.30\pm 0.25}$ \msun\ ($\ms= 10^{10.40\pm 0.25}$ 
\msun), which is around 0.23 (0.30) times $M^\star$, the Schechter fit 
characteristic mass of the overall \gsmf\ of YMB09. For smaller and 
larger masses, \fs\  significantly decreases. 
 
$\bullet$ 
We have compared our results with the few observational inferences 
of the \ms--\mh\ relation for blue (late-type) and red (early-type) galaxies
that exist in the literature. 
Although these studies estimate halo masses using direct techniques 
(weak lensing and galaxy satellite kinematics), they are still 
limited by the stacking approach they need to apply (due to the low signal-to-noise ratio 
of individual galaxies) and by the large uncertainties owing to the unknown systematics. 
The overall differences among the different studies (including ours)
amount up to factors 2-4 at a given mass (these factors being much smaller
at other masses) for most methods (Fig. \ref{comp}). 
For blue galaxies, all methods agree reasonably well for low masses 
($\mh\lesssim 3\times 10^{12}$ \msun), but at higher masses, our inference 
implies larger halos for a given \ms\ than the results from direct techniques. 
For red galaxies, at high masses ($\mh\grtsim 3\times 10^{12}$ \msun), all 
methods agree reasonably well, but at lower masses, the satellite kinematics
technique produces halo masses, for a given \ms, larger than those obtained
by other methods. 

$\bullet$
According to our results, for $\mh\lesssim 10^{11.3}$ \msun, 
the intrinsic scatter of the \ms--\mh\ relation should slightly anti-correlate 
with galaxy color (for a fixed \mh, the bluer the galaxy, the higher its \ms), 
while for more massive systems, the correlation should be direct (for a fixed \mh, 
the redder the galaxy, the higher its \ms). For massive blue 
galaxies in order  to had have higher higher \fs\ values
than the red ones as the results from direct techniques suggest, the \HMF\ halos 
hosting blue (red) galaxies should be even steeper (shallower) 
than what we have proposed here; this seems too exaggerated. 

$\bullet$ 
The maximum baryon mass fraction of blue and red galaxies are 
$\fb=0.028_{-0.011}^{+0.018}$ and $\fb=0.034^{+0.025}_{-0.014}$,
respectively, much smaller than $f_U=0.167$ in both cases, and these maxima
are attained at $\mh\approx 10^{12}$ \msun.  
At large masses \fb\ decreases approximately as $\fb\propto \mh^{-0.5} (\mb^{-0.8})$
for blue galaxies and as $\fb\propto \mh^{-0.6} (\mb^{-1.5})$ for red galaxies, in such a 
way that from $\mh\approx 5\times 10^{12}$ \msun, blue galaxies have on average slightly 
larger values of \fb\ than red ones. At low masses, the \fb\ of red
galaxies strongly decreases as the mass is smaller $\fb\propto  \mh^{2.9} (\mb^{0.8})$, while 
for blue galaxies, due to the increasing gas fractions as smaller is the mass, \fb\ 
decreases slower than \fs, as $\fb\propto  \mh^{0.7}$ ($\propto \mb^{0.4}$).

The AM technique has been revealed as a relatively simple but powerful method
for connecting empirically galaxies to dark halos. Here we extended this 
technique towards inferences for the blue and red galaxy sub-populations
separately. By introducing a minimum of assumptions --otherwise the method
becomes close to a semi-analytical model-- we have found that the stellar
and baryon mass--halo mass relations of blue and red galaxies do not 
differ significantly among them and from the overall ones. The maximum
differences are around the peak of these relations, $\mh\approx 10^{12}$ \msun,
and are consistent qualitatively with the inference that the galaxies in these halos
are transiting from an active to a quiescent regime  of \ms\ growth (FA10). Those that
transited recently is because they had an efficient process of gas consumption
into stars and further cessation of \ms\ growth; therefore, should be redder
and with higher \fs\ values than those that still did not transit. For larger and
lower masses than  $\mh\approx 10^{12}$ \msun, the differences decrease 
and even invert, something that is also consistent with the inferences by FA10, 
based on the semi-empirical determinations of the evolution of the overall 
\ms--\mh\ relation.

\section*{ Acknowledgments}
We thank the Referee for his/her useful comments and suggestions.
We are grateful to Dr. S. More for sending us in electronic form their
data plotted in Fig. \ref{comp}.  A.R-P. acknowledges a graduate student 
fellowship provided by CONACyT. 
We thank PAPIIT-UNAM grant IN114509 and CONACyT grant 60354 
for partial funding, as well as a bilateral DFG-CONACyT
grant through which we got access to results from N-body numerical
simulations performed by Dr. S. Gottloeber.

\end{document}